\documentclass[preprint,showpacs,preprintnumbers,amsmath,amssymb,natbib]{revtex4}

\usepackage{graphicx}
\usepackage{epsfig}		
\usepackage{dcolumn}
\usepackage{bm}

\begin{document}

\title{Accretion of non-minimally coupled generalized Chaplygin gas into black holes}

\author{M. G. Rodrigues}
\email{manuela.rodrigues@ufabc.edu.br}
\affiliation{Centro de Ci\^{e}ncias Naturais e Humanas, Universidade Federal do ABC, Rua Santa Ad\'{e}lia, 166, 09.210-170, Santo Andr\'{e}, SP, Brasil}
\altaffiliation{Also at Instituto de F\'{\i}sica Gleb Wataghin, UNICAMP, PO Box 6165, 13083-970, Campinas, SP, Brasil}
\author{A. E. Bernardini}
\email{alexeb@ufscar.br}
\affiliation{Departamento de F\'{\i}sica, Universidade Federal de S\~ao Carlos, PO Box 676, 13565-905, S\~ao Carlos, SP, Brasil}

\date{\today}

\begin{abstract}
The mass evolution of Schwarzschild black holes by the absorption of scalar fields is investigated in the scenario of the generalized Chaplygin gas (GCG).
The GCG works as a unification picture of dark matter plus dark energy that naturally accelerates the expansion of the Universe.
Through elements of the quasi-stationary approach, we consider the mass evolution of Schwarzschild black holes accreted by non-minimally coupled cosmological scalar fields reproducing the dynamics of the GCG.
As a scalar field non-minimally coupled to the metrics, such an exotic content has been interconnected with accreting black holes.
The black hole increasing masses by the absorption of the gas reflects some consistence of the accretion mechanism with the hypothesis of the primordial origin of supermassive black holes.
Our results effectively show that the non-minimal coupling with the GCG dark sector accelerates the increasing of black hole masses. 
Meanwhile some exotic features can also be depicted for specific ranges of the non-minimal coupling in which the GCG dynamics is substantially modified.
\end{abstract}

\maketitle\thispagestyle{plain}

Accretion of matter by massive compact objects has been currently investigated in the background scenario of the general relativity \cite{bondi}.
Such accretion processes are candidates to mechanisms of formation of supermassive black holes (SMBH) in the center of most active galaxies \cite{moffat}.
In particular, it should follow some analogies with the process proposed by Salpeter and Zeldovich where galaxies and quasars could get some of their energy from processes of accretion \cite{merritt}.

Primordial black holes (PBH) are suggested as those created at the end of inflationary period.
Among several physical processes which prescribe their production, one could suppose some kind of phase transition to the radiation era involving topological defects.
Black holes with $M\leq 10^{15}g$ should not be observed today, as they may have evaporated by Hawking radiation \cite{hawking} and contribute to the flux of cosmic rays.
Those with $M\geq 10^{15}g$ could produce some observable effects at the present epoch \cite{schroedter}, in particular, the effect of gravitational microlensing, or more generically, some significant contribution to the amount of dark matter \cite{carr3}.
When these PBH are formed with sufficiently large mass so that they do not evaporate completely, in case of matter and energy being absorbed in large quantities, these objects should be related to the supermassive black holes observed today \cite{SMBH}.

These theories are all conceived in a scenario where observational evidences strongly suggest that the Universe is dominated by some dark energy component and evolves through a phase of accelerated expansion \cite{Expansion}.
In such a context, the fine-tuning of the vacuum energy and the so-called coincidence problem are still open issues for which the standard $\Lambda CDM$ model \cite{constant} does not provide a complete explanation.
Although the quintessence models \cite{saa,quintessence} are in agreement with CMB data and provide a reasonable explanation to the coincidence problem, the corresponding equation of state, $\omega_{\phi} = p(\phi) /\rho(\phi)$, usually introduces some illogical constraints as, for instance, $-1 \leq\omega_{\phi}\leq 0$.
Another class of dark energy model is resumed by the generalized Chaplygin gas (GCG) \cite{Kam02,Bil02,Ber02}, which is derived from the generalized model of D-branes.
It synthesizes a unified dark sector through an exotic equation of state for dark matter and dark energy.
The GCG is particularly relevant in this respect as it is recognized by observational constraints from CMB \cite{Ber03}, supernova \cite{Sup1,Ber04,Ber05}, gravitational lensing surveys \cite{Ber03B}, and gamma ray bursts \cite{Ber06B}.
Moreover, it has been shown that the GCG model can be accommodated within the usual formation mechanisms for large scale structures \cite{Kam02,Ber02,Ber04}.

Our simplifying hypothesis for the dark sector therefore sets a Universe completely filled by the GCG underlying scalar field, which yields, through its gravitational impact, a viable unified scheme to mimic the presence of both dark energy and dark matter.

The subjacent analysis considers the possibility of accretion of matter by massive black holes from minimal and non-minimal couplings to a real scalar field which drives the dynamics of the GCG \cite{Kam02,Ber04}.
The scalar field pervading all the Universe is assumed as an additional channel for black hole mass accretion.

However, one should notice that the study of growing black hole masses in the presence of scalar field couplings to the metrics has been initiated before the discovery of the late time acceleration of the Universe and, thus, before the proposal of any dark energy accelerated expansion mechanism.
Once inside the black hole horizon, this scalar field would be absorbed and produce the black hole mass increasing.
The growing mass during the life of a PBH is something relevant and cannot be overlooked.
We are mainly interested in modeling the black hole mass accretion by the GCG through the hypothesis of considering its non-minimal coupling to the metrics.

In the above-mentioned context, the GCG model is characterized by an exotic equation of state \cite{Ber02,Ber03} given by,
\begin{equation}
p=-\frac{A}{\rho^{\alpha}},
\label{eq:pch}
\end{equation}
which, for instance, can be deduced from a generalized Born-Infeld action \cite{Ber02}.

By solving the unperturbed equation for energy conservation in case of the equation of state introduced above, one obtains through a straightforward mathematical manipulations \cite{Ber02}
\begin{equation}
\rho_{ch}=\rho_{0}\left[A_{s}+\frac{\left(1-A_{s}\right)}{a^{3\left(1+\alpha\right)}} \right]^{\frac{1}{1+\alpha}},
\label{eq:ch}
\end{equation}
where $\rho_{0}$ is the present energy density of the Universe, and $A_{s}=\frac{A}{\rho_{ch0}^{1+\alpha}}$.
This model interpolates the dust dominate phase in the past, where $\rho\propto a^{-3}$ and the de-Sitter phase, $\rho\propto -p$, at late times. This evolution is phenomenologically constrained by $\alpha$ and $A_{s}$ parameters, both positive, and with $0 < \alpha \leq 1$.
Obviously, $\alpha = 0$ corresponds to the $\Lambda$CDM model.
The case $\alpha = 1$ corresponds to the equation of state of the Chaplygin gas scenario \cite{Kam02} and is already ruled out by data \cite{Ber03}.
Notice that for $A_s =0$, GCG always behaves as matter whereas for $A_{s} =1$, it behaves as a cosmological constant.
Hence to use it as a unified candidate for dark matter and dark energy one has to exclude these two possibilities so that $A_s$ must lie in the range $0 < A_{s} < 1$.
In particular, for a flat universe, $\alpha=0$ corresponds to the $\Lambda$CDM model, and the best fit in agreement with the SNe-Ia datas happens for $\alpha=0,999$ and $A=0,79$ \cite{bertolami}.

By assuming that the GCG fluid has a homogeneous time-dependent underlying scalar field \cite{Kam02,Ber04,Bil02,Ber02}, with an appropriate potential, one can define an action in the form of
\begin{equation}
S=\int d^{4}x\sqrt{-g}\left\{F\left( \phi \right)R-\partial_{a}\phi\partial^{a}\phi-2V\left(\phi\right)\right\}.
\label{action}
\end{equation}
where $V\left(\phi\right)$ is the self-interacting potential and $F\left( \phi \right)$ features the coupling to the metrics.
In the above action, units have been chosen so that ${8}\pi{G}=c=\hbar=k=1$ and the metric signature adopted is $(-+++)$.

By evaluating the above action, one obtains 
\begin{equation}
T_{ab}=\partial _{a}\phi \partial _{b}\phi -\frac{g_{ab}}{2}\left( \partial_{c}\phi \partial ^{c}\phi +2V\right)+\nabla_{a}\nabla _{b}F-g_{ab}\Box F,
\label{einstein}
\end{equation}
and the correlated Klein-Gordon equation described by
\begin{equation}
\Box\phi-V'+\frac{1}{2}F'R=0,
\label{eq:kleing}
\end{equation}
where the prime denotes derivation with respect to $\phi$.

For an isotropic, homogeneous and spatially flat universe, one has
\begin{equation}
ds^{2}=-dt^{2}+a^{2}\left( t\right) \left( dx^{2}+dy^{2}+dz^{2}\right),
\end{equation}
and the temporal component of the Einstein equation gives the energy constraint through the following relation,
\begin{equation}
3H\left( FH-F^{\prime }\phi \right) =\frac{\dot{\phi }^{2}}{2}+V\left(\phi \right),
\label{eq:vinculo}\
\end{equation}
where {\em dot} denotes differentiation with respect to the physical time.
The modified Friedmann equation, for the space-space components of Eq.~(\ref{einstein}), can thus be written as
\begin{equation}
-2\left(F+\frac{3}{2}F^{\prime 2}\right) \dot{H}=3\left(F+2F^{\prime 2}\right) H^{2}+\frac{1}{2}\left( 1+F^{\prime \prime }\right)
\dot{\phi }^{2} -V -F^{\prime }\left( V^{\prime }+H\dot{\phi }
\right),
\label{eq:espaciais}
\end{equation}
where $H=\dot{a}/a$ is the Hubble rate.

\textbf{Minimal coupling}

By setting $F(\phi)=1$ one recovers the minimal coupling and the Klein-Gordon equation becomes
\begin{equation}
\overset{..}{\phi}+3H\phi+V'\left(\phi\right) = 0.
\label{eq:kleingordonminimo}
\end{equation}
Assuming that $G_{ab}=T_{ab}$, the energy-momentum tensor is then given by
\begin{equation}
T_{ab}=\partial_{a}\phi\partial_{b}\phi-\frac{g_{ab}}{2}\left(\partial_{c}\phi\partial^{c}\phi+2V\right),
\label{eq:energiamomentominimo}
\end{equation}
from which the temporal and spatial components respectively gives the density and the pressure as $\rho_{\phi}=\frac{\overset{.}{\phi}^{2}}{2} + V\left(\phi\right)$, and $p_{\phi}=\frac{\overset{.}{\phi}^{2}}{2}-V\left(\phi\right)$.
Using Eqs.~(\ref{eq:ch}) and (\ref{eq:pch}), after some simple mathematical manipulations, one obtains the field equation for the GCG as
\begin{equation}
\frac{d\phi(a)}{da}=\frac{1}{a}\sqrt{\frac{1}{\frac{A_{s}a^{3(1+\alpha)}}{1-A_{s}}+1}}.
\label{campoanalitico}
\end{equation}

\textbf{Non-minimal coupling}

Models with $F(\phi)=0$ exhibit a plenty of singularities.
Homogeneous and isotropic solutions passing from the $F(\phi)<0$ to $F(\phi)>0$ domains are indeed extremely unstable \cite{singularities}.
Thus, our study of the non-minimal coupling has been constrained by the solutions where $F(\phi)>0$.

In this case, the density and pressure are defined from the energy momentum tensor, $T_{ab}=(p+\rho)u_{a}u_{b}+pg_{ab}$, so that, $p_{\phi}=- 2\overset{.}{H}-3H^{2}$, and $\rho_{\phi}=3H^{2}$, so that the temporal component of the Einstein equation is thus written as
\begin{equation}
\rho F+3H\dot{\phi}F'=\frac{\dot{\phi}^{2}}{2}+V.
\label{eq:movimento1}
\end{equation}
Assuming the simplifying notation with $\dot{\phi} = x$ and setting $R = \rho - 3p$, the Eq.~(\ref{eq:kleing}) becomes
\begin{equation}
(x\,x'+V')F'+3HF'\,x -\frac{F'^{2}}{2}(\rho-3p)=0.
\label{eq:novaklein}
\end{equation}
Inserting the redefined variables into Eqs.~(\ref{eq:espaciais}) and (\ref{eq:movimento1}), and combining them with Eq.~(\ref{eq:novaklein}) results into
\begin{equation}
(p+\rho)F=(1+F'')x^{2}+F'x(x'-H).
\end{equation}

Turning back to the previous notation, i. e. $x=\dot{\phi}=a\,H\frac{d\phi}{da}$, one stays with
\begin{equation}
(p+\rho)\frac{3F}{a^{2}\rho ^{2}}=(1+F'')(\phi'(a))^{2}+\frac{F'}{2\rho}\phi'(a)\frac{d\rho}{da}+F'\phi''(a)
\label{eq:mov}
\end{equation}
and the potential can be found through Eq.~(\ref{eq:espaciais}) rewritten as
\begin{equation}
F'V'=\rho(F+2F'^{2})-(p+\rho)\left(F+\frac{3F'^{2}}{2}\right)-\frac{a\rho}{3}F'-V+\frac{1}{2}(1+2F'')\frac{a^{2}\rho}{3}(\phi'(a))^{2},
\label{eq:mov2}
\end{equation}
where, from the beginning, we have assumed that the scalar field dependence on the scale factor $a$ reproduces the background of the GCG.

Assuming a perturbative dependence of the coupling to the metrics on the scalar field through $F(\phi) = 1 - 2\xi\phi^{2}$, where $\xi$ is the coupling constant between the scalar field $\phi$ and the Ricci scalar, we have numerically solved the system of differential equations given by Eqs.~(\ref{eq:mov}) and (\ref{eq:mov2}) in terms of the scale factor, $a$.

\begin{figure}
\begin{center}
\includegraphics{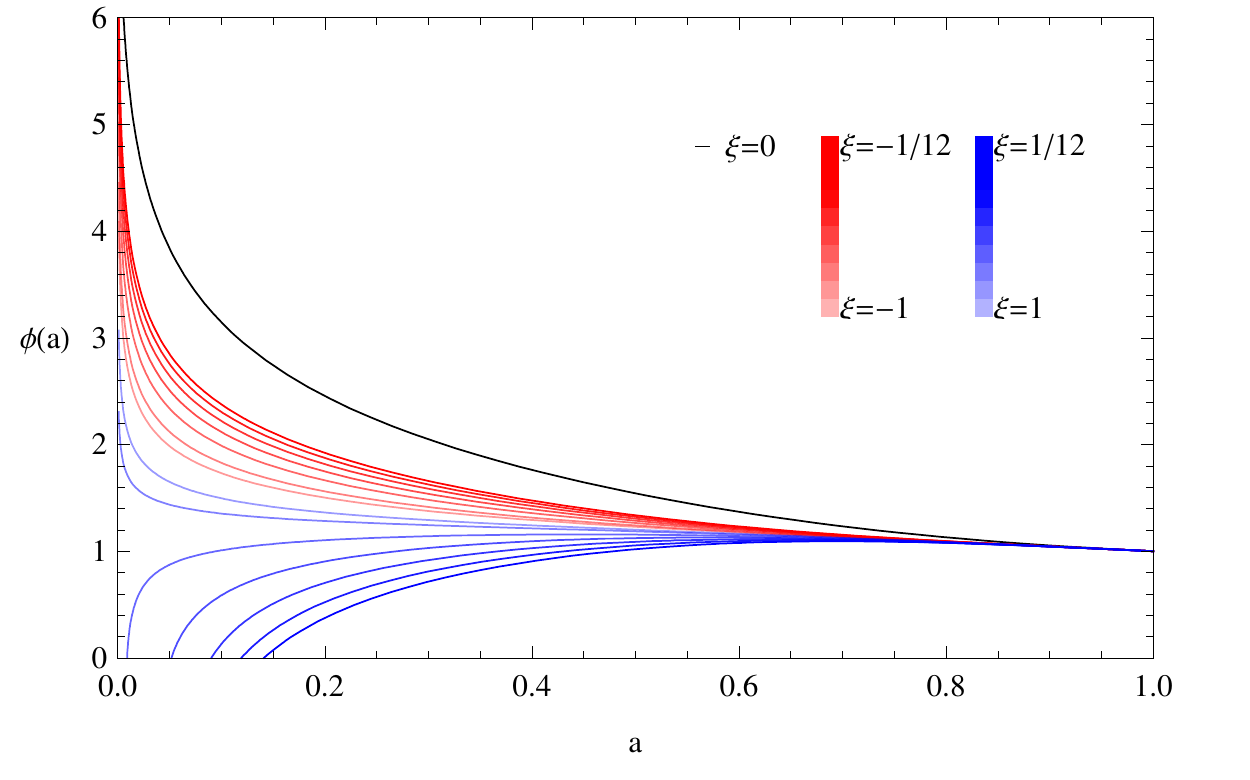}
\end{center}
\caption[Field framework]{\normalsize {Dynamical dependence of the scalar field $\phi$ on the scale parameter $a$, for the GCG model with fixed $\alpha=0,999$ and $A=0,79$, in case of a non-minimal coupling parameterized by $F(\phi)=1-2\xi\phi^{2}$.
Red lines indicate the $\xi < 0$ solutions, blue lines represent the $\xi > 0$ solutions, and the black line corresponds to the minimal coupling solution.}}
\label{phi1}
\end{figure}

In Fig.~\ref{phi1} one can depict the scalar field dependence on $a$ in case of $-1 < \xi < 1$.
Notice that for the solutions where $\xi < 0$ (red lines), the scalar field qualitatively behaves as the minimally coupled solution for the GCG.
Due to the non-linearity of the system of differential equations that we have solved, there is no continuous transition from negative to positive solutions for $\xi$.
One shall notice the same qualitative behavior for values of $\xi \gtrsim 1/3$.
However, for the interval where $0 < \xi \lesssim 1/3$, which includes the conformal coupling, $\xi=1/6$, one has singularities when $a$ tends to zero.
\begin{figure}[h!]
\begin{center}
\includegraphics{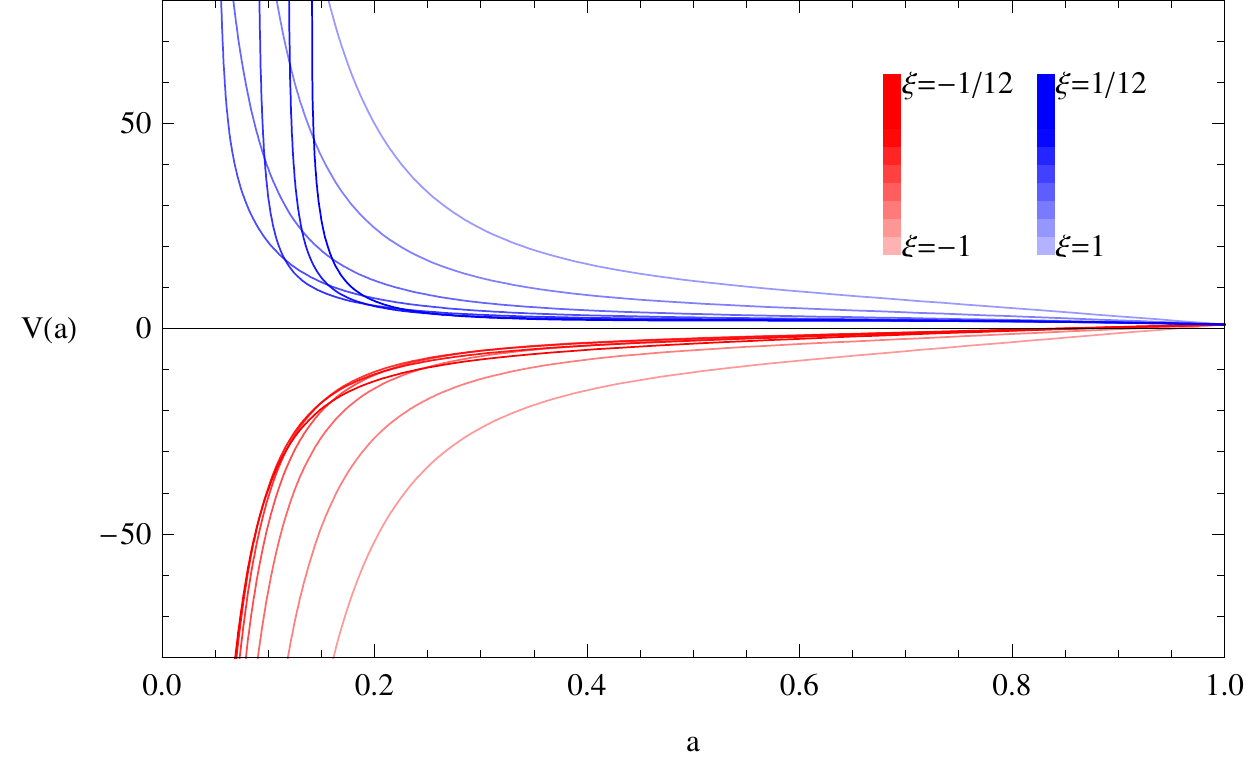}
\end{center}
\caption[Potencial] {\normalsize{The aspect of the potential running through the scale factor, in correspondence with Fig.~\ref{phi1}.
Blue lines represent $\xi > 0$ solutions and red lines represent $\xi < 0$ solutions.}}
\label{potencial1}
\end{figure}

As we have implicitly asked at the beginning of our exposition, does the picture of a Universe described by the GCG is consistent with the models for accretion of mass into black holes?
In a black hole, the gravitational force is so intense that any form of matter or radiation is next to be drawn into it.
In the region of the event horizon of a Schwarzschild black hole, the mass variation has two factors, the amount of matter and energy present at the border, which will be absorbed, and the evaporation by Hawking radiation.
In spite of being studied for the case of ordinary matter, perfect fluid and scalar fields \cite{barrow}, until the discovery of accelerated expansion of the universe, the accretion of matter was not yet correlated to dark energy components.

We have investigated the process of mass accretion and matter evaporation, through the above obtained results for the Chaplygin gas.
In case of a scalar field in the vicinity of a black hole, assuming that on the horizon one has $r = 2M$, all the energy flux would be absorbed. According to the Bondi accretion \cite{bondi}, for which the field is stationary and spherically and symmetrically absorbed, the black hole mass increasing would be given by
\begin{equation}
\dot{M}=\oint_{r=2M}d\Omega r^{2}T_{0}^{r}= 4\pi r^{2}T_{0}^{r},
\end{equation}

The amount of matter/energy evaporated by Hawking radiation, in this case, is given by
\begin{equation}
\overset{.}{M}=-\frac{\beta}{M^{2}},
\label{eq:evaporacao}
\end{equation}
where $\beta$ is a model dependent constant \cite{gama}.
At first glance, and for simplicity, we have considered non-rotating black holes in the Einstein frame.
We have assumed the Schwarzschild metric as
\begin{equation}
ds^{2}=-\left(1-\frac{2M}{r}\right)dt^{2}+\left(1-\frac{2M}{r}\right)^{-1}dr^{2}+r^{2}\left(d\theta^{2} + \sin{(\theta)}^{2}\,d\varphi^{2}\right),
\label{eq:metric}
\end{equation}
from which the spherically symmetric version of the Klein-Gordon equation obtained from Eq.~(\ref{action}) can be written as
\begin{equation}
-\frac{\partial^{2}\phi}{\partial t^{2}}+\frac{1}{r^{2}}\left(1-\frac{2M}{r}\right)\frac{\partial}{\partial r}\left[r^{2}\left(1-\frac{2M}{r}\right)\frac{\partial \phi}{\partial r}\right]=\left(1-\frac{2M}{r}\right)V'(\phi).
\label{eq:klein}
\end{equation}

At the vicinity of a black hole, the cosmological scalar field could generate a gravitational field much weaker than the gravitational field black hole.
Although one can approximates the metrics by the Schwarzschild one in this region, a relation between the scalar field that was approximated from the infinity to the black hole horizon, namely the field absorbed by the black hole, is demanded.
By assuming the quasi-stationary solution \cite{frolov} for a slow-variation of the field in the nearby black hole, one should have
\begin{equation}
\phi\left(r,t\right)\approx\phi_{c}\left[v-r+2M\log\left(\frac{2M}{r}\right)\right].
\label{eq:quaseestacionaria}
\end{equation}
Substituting the above solution into Eq.~(\ref{eq:klein}), it is possible to extend the validity of the approximation to $\overset{..}{\phi}_{c}\approx0$ and $V'(\phi_{c})\approx0$ \cite{rodrigues}.
The field on the event horizon can thus be parameterized by the same cosmological field approximated from the infinity with some time delay,
\begin{equation}
\phi_{c}\left(t\right)\approx \phi_{\infty}+\dot{\phi}_{\infty}t,
\end{equation}
where $\phi_{\infty}$ and $\dot{\phi}_{\infty}$ are constants obtained from the temporal solution of Eqs.~(\ref{eq:kleing}), (\ref{eq:vinculo}) and (\ref{eq:espaciais}).
The scale factor can be depicted from Eq.~$\rho_{\phi}=\rho_{ch}$, so that
\begin{equation}
\dot{a}=\frac{a\sqrt{3\rho_{ch0}\left[A_{s}+\frac{\left(1-A_{s}\right)}{a^{3\left(1+\alpha\right)}} \right]^{\frac{1}{1+\alpha}}}}{\sqrt{3}}.
\end{equation}
Using this solution into Eq.~(\ref{einstein}) one obtains
\begin{equation}
T_{0}^{r}= \frac{2M}{r^{2}}\left[2M\left(1+F''\right)\overset{.}{\phi}^{2}_{\infty}  -\frac{F'}{2}\overset{.}{\phi}_{\infty}\right],
\label{eq:temporadial}
\end{equation}
and the mass distribution in dependence on the scale factor becomes
\begin{equation}
\dot{M}=16\pi M^{2}\left[\left(1+F''\right)\dot{\phi_\infty}^{2}-\frac{F'}{4 M}\dot{\phi}\right] -\frac{\alpha}{M^{2}}.
\end{equation}

\begin{figure}
\begin{center}
\includegraphics{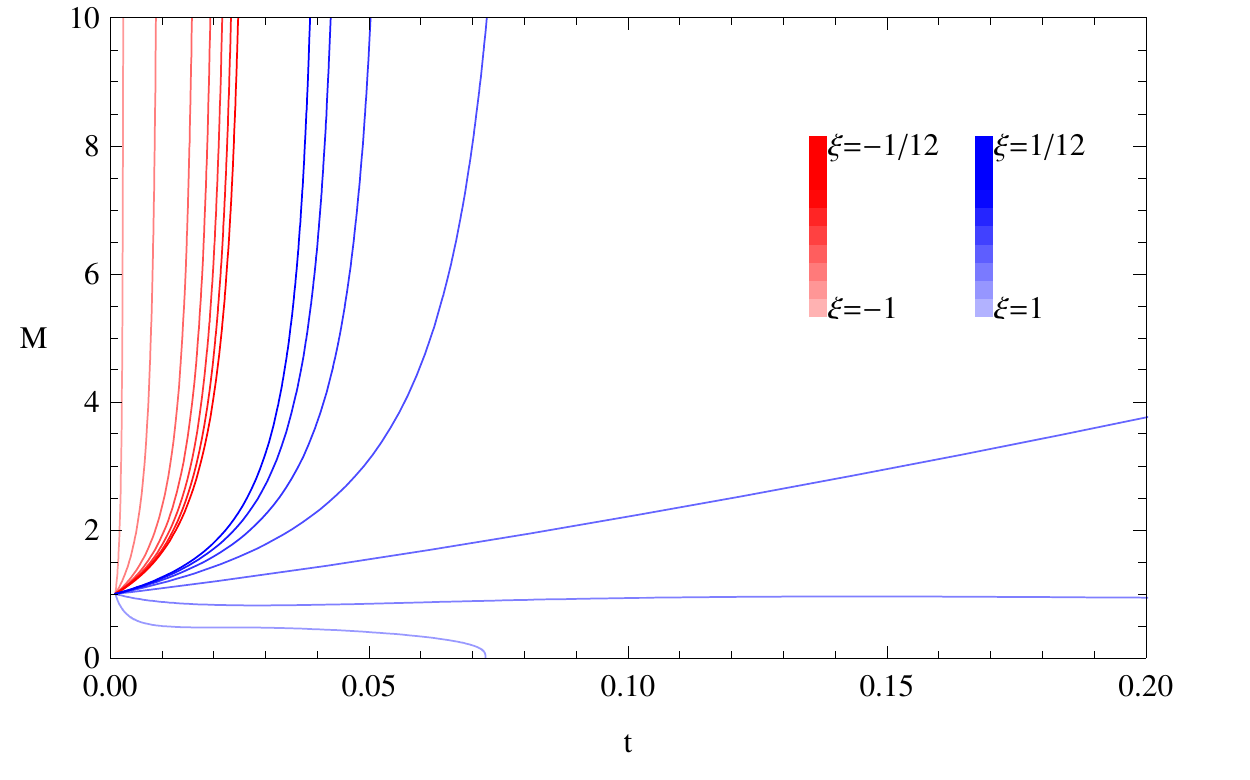}
\end{center}
\caption {\normalsize{Mass evolution of a black hole on the vicinity of the GCG.
We have computed the effects for the coupling parameter in the interval $-1 < \xi < 1$.
The red lines show $-1 < \xi < 0$ solutions and the blue ones show $0 < \xi < 1$ solutions.
The mass, $M$, is in units of $M_{0}$ and the time, $t$, is in units of the Hubble time.}}
\label{naomin}
\end{figure}

For a minimal coupling, and considering the Hawking radiation, one thus get
\begin{equation}
\dot{M}=16\pi M^{2}\dot{\phi_\infty}^{2}-\frac{\beta}{M^{2}}.
\end{equation}
Fig.~\ref{naomin} shows the mass distribution considering the Hawking radiation for a non-minimally coupled scalar field, using $1 < \xi < -1$ for $M_{0} = 1$, as a reference value.
For $1/3 < \xi < -1$ the mass increases faster and indefinitely, but for $\xi > 1/3$, one can notice some instability effect so that a decreasing mass behaviour is identified.

The mass evolution and the scalar field dependence on the cosmological scale parameter, $a$, are semi-analytically obtained for a perturbative non-minimal coupling of the GCG underlying scalar field with the Ricci scalar of the metric.
It reveals some qualitatively distinct behaviors for the black hole mass, $M(t)$, and for the scalar field, $\phi(a)$, when compared with the results for the minimal coupling case. 
For black hole masses smaller than a certain critical value assumed by the non-minimal coupling parameter, $\xi > 1/3$, the accretion of the scalar field can lead to an exotic {\em mass decreasing effect}.

By observing the correspondence between the solutions depicted from Figs.~\ref{phi1} and \ref{naomin}, respectively for $\phi(a)$ and $M(t)$, one can notice that the curves described by red lines for $\phi(a)$ follows the same qualitative behavior of the free GCG solution described by a black line.  

When the competition between dark matter and dark energy cosmological phases of the unifying picture of the GCG is dominated by dark matter, the black hole mass evolution reproduces the accelerated mass increasing (red lines) as usual for the non-minimal coupling involving ordinary matter fields.
Some previous issues involving modified $f(R)$ gravity models can also reproduce the same effects.
Otherwise, the dark energy dominated phase of the GCG naturally accelerates the expansion of the Universe.
For a specific range of the parameter $\xi$, the inclusion of the non-minimal coupling into the equations anticipates the dark energy phase of the scalar field as one can notice through the blue curves describing the dynamics of $phi(a)$.
For $d\phi/da \sim 0$ at Fig.~\ref{phi1}, one has the anticipation of the cosmological constant phase, and for $d\phi/da > 0$ one has a kind of phantom dark energy effect (with an exotically negative kinetic energy term that leads to the violation of the strong energy condition).
The peculiar {\em mass decreasing effect} quoted above follows this kind of phantom dark energy dominated phase which takes the place of the suppressed dark matter dominated phase.
Even if no phantom field is involved, the non-minimal coupling with the GCG unifying fluid provides such an effective picture for this. 

In more general lines, the mass evolution of black holes is governed by two simultaneously competing mechanisms.
First of all, the Hawking radiation decreases the black hole mass due to the emission of thermal radiation.
In parallel, the accretion of the surrounding available matter and energy usually increases the black hole mass.
Possible evidences of primordial black holes at present shall depend on the balance of these processes and, as one could notice from our analysis, on the nature of the Universe's underlying scalar field, and on its couplings. 
The unexpected hypothesis of black hole mass decreasing because of the accretion of exotic (phantom) dark energy could change qualitatively the evolution of any black hole, implying, occasionally, into novel observational aspects for both astrophysical and PBH's.

Reporting about the quasi-stationary approach, we have considered the mass evolution of Schwarzschild black holes in the presence of a non-minimally coupled cosmological scalar field that analytically resumes a unifying prescription for dark matter and dark energy of the Universe.
Some imprints of such a unifying formulation onto the evolution of super-massive black holes and, reciprocally, onto the cosmological behavior of the scalar field have been discussed.

Such results summarize our investigation of some generic possibilities of non-minimal coupling of a dark sector scalar field to the Ricci scalar.
As we have supposed, it could result into some changing on the dark sector dynamics, as well as, on novel ingredients to the mechanism of accretion of matter into black holes.
We have found that for a non-minimal coupling drven by $\xi > 0$ some unexplained instabilities are present.
It reinforces that only modified couplings parameterized by $F(\phi) > 0$ could be used.
The analysis of the mass distribution of a black hole embedded by the GCG dynamical field allows for connecting PBH's to standard model black holes, in case of GCG absorption.
Such result is in complete agreement with those found in the previous analysis \cite{SMBH, rodrigues, chaplygin2}, where the mass of the black hole reproduces the increasing behavior predicted by the more fundamental models.
In addition, it also reproduces some previous arguments in favor of the cosmology ruled by the GCG model.

\begin{acknowledgments}
A. E. B. is thankful to the financial support from the Brazilian Agencies FAPESP (grant 12/03561-0) and CNPq (grant 300233/2010-8).
M. G. R. is also thankful to CNPq (PD grant 150034/2010-5).
\end{acknowledgments}

\end{document}